\documentstyle[times,balanced,epsf,prl,aps]{revtex}

\begin{document}

\draft

\title{Nonlinear Dynamics of the Perceived Pitch of Complex Sounds}

\author{
Julyan H. E. Cartwright$^{1,}$\cite{jemail}, 
Diego L. Gonz\'alez$^{2,}$\cite{demail}, and 
Oreste Piro$^{3,}$\cite{oemail}
} 

\address{
$^1$Instituto Andaluz de Ciencias de la Tierra, IACT (CSIC-UGR),
E-18071 Granada, Spain. \\
$^2$Istituto Lamel, CNR, I-40129 Bologna, Italy \\
$^3$Institut Mediterrani d'Estudis Avan\c{c}ats, IMEDEA (CSIC--UIB),
E-07071 Palma de Mallorca, Spain \\ 
}

\date{Physical Review Letters, {\bf 82}, 5389--5392, 1999}

\maketitle

\begin{abstract}
We apply results from nonlinear dynamics to an old problem in acoustical 
physics: the mechanism of the perception of the pitch of sounds, 
especially the sounds known as complex tones that are important for music and 
speech intelligibility.
\end{abstract}


\begin{twocolumns}

The pitch of a sound is where we perceive it to lie on a musical scale. Like
all sensations, pitch is a subjective quantity related to physical attributes 
of the stimulus, in this case mainly its component frequencies. For a pure tone
with a single frequency component, the relation is monotonic and enables us to
adopt an operational definition of pitch in terms of frequency: the pitch of an
arbitrary sound is given by the frequency of a pure tone of  the same pitch.
The scientific investigation of pitch has a long history dating back to
Pythagoras,  but the origin of the pitch of complex tones with several
frequency components  is still not well understood. The first perceptual
theories considered pitch to arise at a peripheral level in the auditory system
\cite{ohm,seebeck,helmholtz,schouten,plomp}; more recently it has been thought
that central nervous system processing is necessary
\cite{goldstein,wightman,terhardt}. However, the experimental evidence is that
this processing is carried out before the primary auditory cortex
\cite{pantev}. The latest models integrate neural and peripheral
processing \cite{meddis,cohen2}. Here we develop a nonlinear theory of pitch
perception for complex tones that describes experimental results on the pitch
perception of complex sounds at least as well as do current models, and that
removes the need for extensive processing at higher levels of the auditory
system.  

A key phenomenon in pitch perception is known as the problem of the {\em
missing fundamental}, {\em virtual pitch}, or {\em residue perception} 
\cite{deboer}, and consists of the perception of a pitch that cannot be mapped
to any frequency component of the stimulus. Suppose that a periodic tone such 
as that shown in Fig.~\ref{fig_complex}a is presented to the ear. Its pitch is
perceived to be that of a pure tone at the frequency of the fundamental. The
number of higher harmonics and their relative amplitudes give timbral
characteristics to the sound, which allow one to distinguish a trumpet from a
violin playing the same musical note. Now suppose that the fundamental and some
of the first few higher harmonics are removed (Fig.~\ref{fig_complex}b). 
Although the timbre changes, the pitch of the tone remains unchanged and equal 
to the missing fundamental; this is residue perception. 

The first physical theory for the residue is due to von Helmholtz 
\cite{helmholtz}, who attributed it to the generation of difference 
combination tones in the nonlinearities of the ear. A passive nonlinearity fed
by two sources with frequencies $\omega_1$ and $\omega_2$ generates 
combination tones of frequency $\omega_{C}$, which are nontrivial solutions of 
the equation $p\omega_1+q\omega_2+\omega_{C}=0$, where $p$ and $q$ are
integers. For a harmonic complex tone (Fig.~\ref{fig_complex}b), the difference
combination tone $\omega_C=\omega_2-\omega_1$ (i.e., $p=1$, $q=-1$) between
two  successive partials has the frequency of the missing fundamental
$\omega_0$, that is $\omega_C=(k+1)\omega_0-k\omega_0=\omega_0$. However, a
crucial  experiment seriously challenged nonlinear theories of the residue.
Schouten et al.\ \cite{schouten} demonstrated that the behavior of the residue
cannot be described by a difference combination tone: if we shift all the
partials by the same amount $\Delta\omega$ (Fig.~\ref{fig_complex}c), the
difference combination tone remains unchanged, and the same should thus be true
of the residue. Instead it is found that the perceived pitch also shifts,
showing a linear dependence on $\Delta\omega$ (Fig.~\ref{fig_complex}d). This
phenomenon is known as the first pitch shift effect, and has been accurately
measured in many experiments (psychoacoustic experiments on pitch can attain an
accuracy of 0.2\% \cite{hartmann}). A first attempt to model qualitatively the
behavior of the pitch shift shows that the slopes of the lines in
Fig.~\ref{fig_complex}d depend roughly on the inverse of the harmonic number
$k+1$ of the central partial of a three-component complex tone. However, for
small $k$ at  least, the change in slope is slightly but consistently larger
than this, but smaller if we replace $k+1$ by $k$. This behavior is known as
the second pitch shift effect. Also, an enlargement of the spacing between
partials while maintaining fixed the central frequency produces a decrease in
the residue pitch. As this anomalous behavior seems to be correlated with the
second pitch shift effect, it is usually included within it \cite{schouten}.
Pitch-shift experiments, and others that demonstrated that the residue is
elicited dichotically (part of the stimulus exciting one ear and the rest of
the stimulus the other, controlateral, ear) and not just monotically and
diotically (all of the stimulus exciting one or both ears) led to the
abandonment of peripheral (periodicity \cite{seebeck,schouten} and place
\cite{ohm,helmholtz}) theories and to the development of theories that
considered the pitch of complex tones to be a result of central nervous system
processing \cite{goldstein,wightman,terhardt}; thence to integrated neural and
peripheral models \cite{meddis,cohen2}.

We demonstrate below that the crucial pitch-shift experiment of Schouten et
al.\ can be accurately described in terms of generic attractors of nonlinear 
dynamical systems, such that a theory of pitch perception for complex tones can
be constructed without the need to resort to extensive central processing. We 
model the auditory system as a generic nonlinear forced oscillator,
and identify experimental data with structurally stable behavior of this class
of dynamical system. We emphasize that since we are interested in universal
behavior, the results we obtain are not dependent on the construction of a
particular model, but rather represent the behavior of any dynamical system 
of this type. 

\begin{figure}
\begin{center}
\def\epsfsize#1#2{0.38\textwidth}
\leavevmode
\epsffile{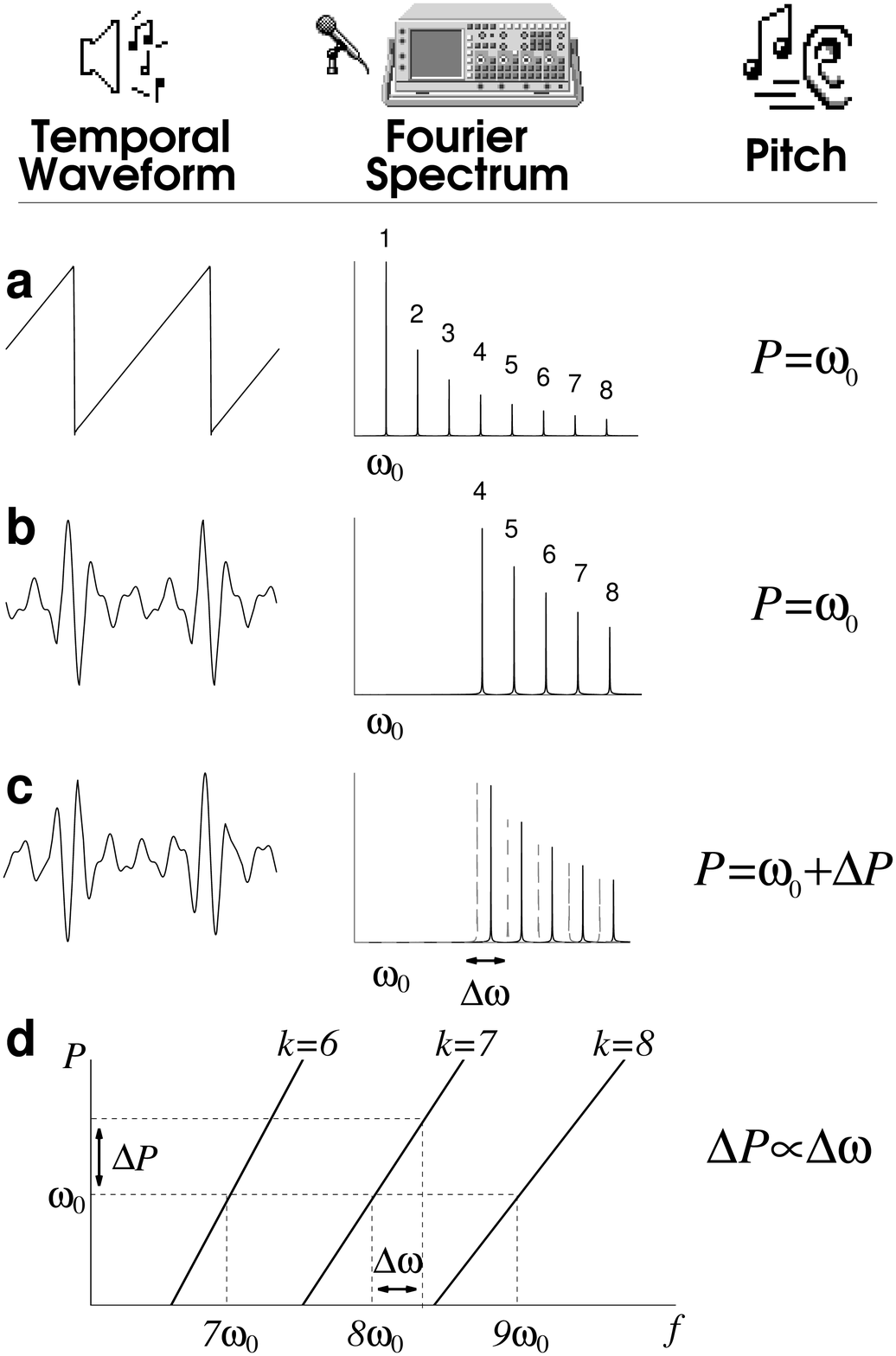}
\end{center}
\caption{\label{fig_complex}
Fourier spectra and pitch of complex tones. Whereas pure tones have a
sinusoidal waveform corresponding to a single frequency, almost all musical 
sounds are complex tones that consist of a lowest frequency component, or 
fundamental, together with higher frequency overtones. The fundamental plus 
overtones are together collectively called partials.
{\bf a} A harmonic complex tone. The overtones are successive integer 
multiples $k=2,3,4\ldots$ of the fundamental $\omega_0$ that determines the 
pitch. The partials of a harmonic complex tone are termed harmonics.
{\bf b} Another harmonic complex tone. The fundamental and the first few higher
harmonics have been removed. The pitch remains the same and equal to the
missing fundamental. This pitch is known as virtual or residue pitch.
{\bf c} An anharmonic complex tone. The partials, which are no longer harmonics,
are obtained by a uniform shift $\Delta\omega$ of the previous harmonic case 
(shown dashed). Although the difference combination tones between successive 
partials remain unchanged and equal to the missing fundamental, the pitch 
shifts by a quantity $\Delta P$ that depends linearly on $\Delta\omega$.
{\bf d} Pitch shift. Pitch as a function of the central frequency 
$f=(k+1)\omega_0+\Delta\omega$ 
of a three-component complex tone $\{k\omega_0+\Delta\omega, 
(k+1)\omega_0+\Delta\omega, (k+2)\omega_0+\Delta\omega\}$. The pitch-shift 
effect is shown here for $k=6$, $7$, and $8$. Three-component complex
tones are often used in pitch experiments because they elicit a clear residue 
sensation and can easily be obtained by amplitude modulation of a pure tone of 
frequency $f$ with another pure tone of frequency $\omega_0$. When $\omega_0$ 
and $f$ are rationally related, $\Delta\omega=0$, and the three frequencies are 
successive multiples of some missing fundamental. At this point $\Delta P=0$,
and the pitch is $\omega_0$, coincident with the frequency of the missing 
fundamental.
}
\end{figure}

Since complex tones can be decomposed as a series of partials --- a sum of
purely sinusoidal components --- and residue perception is elicited with at
least two of these, we search for suitable attractors in the class of
two-frequency quasiperiodically forced oscillators. Quasiperiodically forced
oscillators show a great variety of qualitative behavior that falls into the
three categories of periodic attractors, quasiperiodic attractors, and strange
attractors (both chaotic and nonchaotic). We propose that the residue behavior
we seek to explain is a resonant response to forcing the auditory system
quasiperiodically. From stability arguments 
\cite{diegophd,ourosc,3freq,3freqpaper}
we single out a particular type of two-frequency quasiperiodic attractor,
which we term a three-frequency resonance, as the natural candidate for
modelling the residue. Three-frequency resonances are given by the nontrivial
solutions of the equation $p\omega_1+q\omega_2+r\omega_{R}=0$, where $p$, $q$,
and $r$ are integers, $\omega_1$ and $\omega_2$ are the forcing frequencies,
and $\omega_R$ is the resonant response, and can be written compactly in the
form $(p,q,r)$. Notice that combination tones are three-frequency resonances of
the restricted class $(p,q,1)$. This is the only type of response possible from
a passive nonlinearity, whereas a dynamical system such as a forced oscillator
is an active nonlinearity with at least one intrinsic frequency, and can exhibit
the full panoply of three-frequency resonances, which include subharmonics of
combination tones. The investigation of three-frequency systems is a young area
of research, and there is not yet any consistency in the nomenclature of these
resonances in the scientific literature: as well as three-frequency resonances,
they are also called weak resonances or partial mode lockings; see Baesens et
al.\ \cite{baesens} and references therein. It is known that
three-frequency resonances form an extensive web in the parameter space of
a dynamical system. In particular, between any two parent resonances 
$(p_1,q_1,r_1)$ and $(p_2,q_2,r_2)$ lies the daughter resonance  
$(p_1+p_2,q_1+q_2,r_1+r_2)$ on the straight line in parameter space connecting 
the parents. 

For pitch-shift experiments, the vicinity of the external frequencies to
successive multiples of some missing fundamental ensures that $(k+1)/k$ is a
good rational approximation to their frequency ratio. Hence we concentrate on a
small interval around the missing fundamental between the frequencies
$\omega_1/k$ and $\omega_2/(k+1)$, which correspond to the resonances
$(0,-1,k)$ and $(-1,0,k+1)$. We suppose that the residue should be associated
with the largest resonance in this interval. In numerical simulations and 
experiments with electronic oscillators 
\cite{diegophd,ourosc,3freq,3freqpaper}, we have confirmed that 
for small nonlinearity the daughter of these resonances, $(-1,-1,2k+1)$, is
the resonance of greatest width between its parents. Hence we propose that
the three-frequency resonance formed between the two lower-frequency
components of the complex tone and the response frequency
$P=(\omega_1+\omega_2)/(2k+1)$ gives rise to the perceived residue pitch $P$.
Since in pitch-shift experiments the external frequencies are
$\omega_1=k\omega_0+\Delta\omega$ and $\omega_2=(k+1)\omega_0+\Delta\omega$,
the shift of the response frequency with respect to the missing fundamental is
$\Delta P=2\Delta\omega/(2k+1)$. This equation gives a linear dependence of the
shift on the detuning $\Delta\omega$, in agreement with the first pitch shift
effect. 

\begin{figure}
\begin{center}
\def\epsfsize#1#2{0.48\textwidth}
\leavevmode
\epsffile{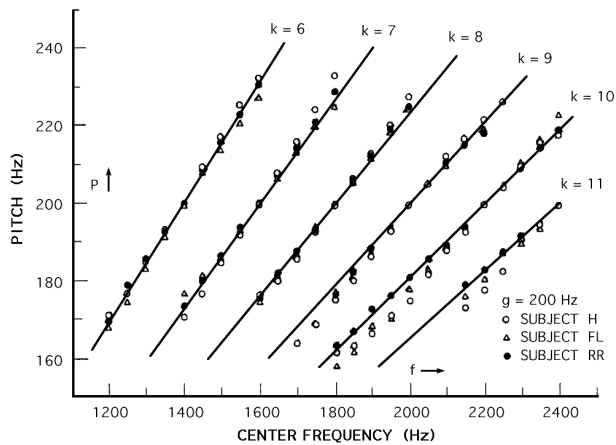}
\end{center}
\caption{\label{fig_pitch_shift2}
Experimental data with our predictions superimposed show pitch as a function of
the central frequency $f=(k+1)\omega_0+\Delta\omega$ for a three-component 
complex tone $\{f-g, f, f+g\}$, for $k=6$, $7$, $8$, $9$, $10$, and $11$. The
component spacing is $g=\omega_0=200$ Hz. Circles, triangles and dots represent
experimental data for three different subjects (from Schouten et al.
\protect\cite{schouten}). The perceived pitch shifts linearly with the detuning
$\Delta\omega$. The pitch shift effect we predict is shown superimposed on the
data as solid lines satisfying the equation $P=g+2(f-(k+1)g)/(2k+1)$. The lines
describe the behavior of the response frequency in a three-frequency 
resonance formed with the two lowest-frequency components of the stimulus. The 
lines agree with the psychophysical data without any fitting parameters.
}
\end{figure}

In Fig.~\ref{fig_pitch_shift2} we have superimposed the behavior of the
corresponding three-frequency resonances on data for the pitch shift for three
different subjects. There is good agreement, which explains both the first
pitch shift effect and the first aspect of the second pitch shift effect,
because the predicted slope is $1/(k+1/2)$, between $1/k$ and $1/(k+1)$. 
The second aspect of the second pitch shift effect can be interpreted as
follows: the term $2\Delta\omega$ in the equation for $\Delta P$ arises from
the two equal contributions $\Delta\omega$ obtained by a uniform shift in the
two forcing frequencies. If, while maintaining $\omega_2$ fixed, we increase the
interval to $\omega_1$ --- we enlarge the spacing between successive partials
--- the first contribution remains constant and equal to $\Delta\omega$ while
the second diminishes, to give a decrease in the response frequency of the
resonance and thus in the residue. 

For small harmonic numbers $k$ our model data fall within experimental error 
bars for most of the pitch-shift experiments reported 
\cite{schouten,patterson,gerson}. For larger $k$ there are systematic
deviations of the residue slopes which become shallower than theoretical
predictions of both our theory, and the central theories. An accepted
explanation for this is that difference combination tones generated as a
consequence of passive nonlinear mixing in the auditory periphery can play the
r\^ole of the lowest frequency components of the stimulus. The same explanation
is also valid for our approach. Our goal here, however, is simply to
demonstrate that physical frequencies other than combination tones can
accurately describe residue behavior in nonlinear terms without fitting
parameters, and can overcome the objections against nonlinear theories raised
by pitch-shift experiments.

Dichotic perception has been another argument against nonlinear theories,
because it has been thought to imply central nervous system processing
\cite{houtsma}. However, subcortical nervous paths, for example through the
superior olivary complex, allow the exchange of information between both ---
left and right --- peripheral auditory systems. Moreover, frequency information
up to 5 kHz is preserved until at least this point in the auditory system
\cite{kiang,johnson2}. These two factors allow frequency components of the
stimulus that arrive at different ears to interact at a subcortical level. This
implies that three-frequency resonances can be generated, not only monotically
but also dichotically, by a mechanism such as that which we are proposing.
Experimental evidence for the existence of multifrequency resonant responses is
given by electrophysiological records in single units of the auditory midbrain
nucleus of the guinea fowl \cite{langner}. Our mechanism is also consistent
with neuromagnetic measurements performed on human subjects showing that pitch
processing of complex tones is carried out before the primary auditory cortex
\cite{pantev}.

Thanks to the residue we can appreciate music in a small radio with negligible 
response at low frequencies; but it is not just an acoustical curiosity. The
residue seems to play an important r\^ole in music perception and speech
intelligibility. The importance of the residue for the perception of musical
sounds has long since been recognized. It has been proposed \cite{terhardt}
that residue perception is at the heart of the fundamental bass of Rameau
\cite{rameau}. As the fundamental bass and its more modern counterparts form a
key element for the melodic and harmonic structuration of musical sounds, a
physical basis for the residue may contribute to the construction of an
objectively grounded theory of music \cite{diegophd}. As for speech
intelligibility, hearing aids that furnish fundamental frequency information
produce better scores in profoundly hearing impaired subjects than simple
amplification \cite{faulkner}. It is clear that a better knowledge of the 
basic mechanisms involved in pitch perception will allow a similar improvement 
in its applications to technology and medicine.

The principle of economy, widespread in nature, suggests that by means of our
proposed mechanism, complex tones may be preprocessed at a subcortical level,
to feed optimized pitch candidates to more central zones of the nervous
system, freeing them of an enormous quantity of calculation. In this way, the
nonlinear behavior of the auditory periphery may greatly contribute to explain
the astonishing real-time analytical capabilities of the auditory system
\cite{pitchpage}.

JHEC and OP acknowledge the financial support of the Spanish Direcci\'on 
General de Investigaci\'on Cient\'\i fica y T\'ecnica, contracts PB94-1167 and 
PB94-1172. We are pleased to thank Egbert de Boer, Manfred Schroeder, and 
Gene Stanley for helpful conversations.

\end{twocolumns}
\end{document}